\documentclass{article}
\def\DR{\rm I\kern-1.45pt\rm R}
\def\DC{\kern2pt {\hbox{\sqi I}}\kern-4.2pt\rm C}

\newcommand{\nn}{\nonumber\\}
\newcommand{\p}[1]{(\ref{#1})}

\newcommand{\bl}{\bar\lambda }

\newcommand{\hL}{{\hat \Lambda}}
\newcommand{\hbL}{{\hat{\overline \Lambda}}}

\newcommand{\bLam}{\overline\Lambda }

\newcommand{\bQ}{{\overline Q}}

\newcommand{\bxi}{{\bar\xi}}
\newcommand{\bpsi}{{\bar\psi}}

\newcommand{\ba}{\begin{array}}
\newcommand{\ea}{\end{array}}
\newcommand{\be}{\begin{equation}}
\newcommand{\ee}{\end{equation}}
\newcommand{\bea}{\begin{eqnarray}}
\newcommand{\eea}{\end{eqnarray}}
\newcommand{\bi}{\begin{itemize}}
\newcommand{\ei}{\end{itemize}}

\newcommand {\bD}{\overline{D}}

\usepackage{amscd,amsmath,amssymb}

\begin{document}
\thispagestyle{empty}
\vspace{2cm}
\begin{flushright}
\end{flushright}
\begin{center}
{\Large\bf N=4 Supersymmetric MICZ-Kepler systems on $S^3$}
\end{center}
\vspace{1cm}

\begin{center}
{\large\bf S.~Bellucci${}^{a}$, S.~Krivonos${}^{b}$, V.~Ohanyan${}^{c}$ }
\end{center}

\begin{center}
${}^a$ {\it INFN-Laboratori Nazionali di Frascati,
Via E. Fermi 40, 00044 Frascati, Italy}

\vspace{0.2cm}

${}^b$ {\it Bogoliubov  Laboratory of Theoretical Physics, JINR, 141980 Dubna,
Russia}

\vspace{0.2cm}
${}^c$ {\it Yerevan State University, A.Manoogian, 1, Yerevan,
375025 Armenia\\
 Yerevan Physics Institute, Alikhanian Brothers St., 2,
Yerevan, 375036, Armenia}

\vspace{1cm}

bellucci@lnf.infn.it, krivonos@theor.jinr.ru, ohanyan@yerphi.am

\end{center}
\vspace{2cm}

\begin{abstract}
Proceeding from the superfield action for $N=4, d=1$ nonlinear
supermultiplet, equipped with the most general potential term, we
find the action describing a charged particle on the sphere $S^3$
in the field of $n$ fixed Dirac dyons. We construct the
supercharges and Hamiltonian and analyze some particulary
interesting potentials corresponding to the $N=4$ supersymmetric
extension of the integrable one- and two-center
McIntosh--Cisneros--Zwanziger--Kepler (MICZ-Kepler) systems on
$S^3$.

\end{abstract}

\newpage
\setcounter{page}{1}
\setcounter{equation}{0}
\section{Introduction}
 The McIntosh--Cisneros--Zwanziger--Kepler(MICZ--Kepler) system is
 the integrable mechanical model which generalizes the Kepler
 (Coulomb)problem for the situation, when the conventional Coulomb
 center is replaced by the Dirac dyon, i.e. a particle carrying both
 electric and magnetic charges. The main feature of this system
 consists in the additional centrifugal potential $U_{MICZ}(r)=\frac{s^2}{2 m r^2}$
 term\footnote{$s=eg$ is the so--called monopole number, $e$ and $m$ are the electric charge and
 the mass of the probe particle, $g$ is the magnetic charge of dyon.}  which
 appears in the Hamiltonian due to monopole-like nature of the forced
 center \cite{MICZ}. The properties of the MICZ--Kepler system are rather similar
 to the ordinary Coulomb one. For instance, beside the conserved angular momentum, the system has
 another integral of motion which is the perfect analog of the
 Laplace--Runge-Lenz vector. At the classical level, trajectories in
 the MICZ--Kepler system have the same shape as in the underlying
 Coulomb one, but in contrast to the latter case, the orbital
 plane is not always orthogonal to the angular momentum. Being
 quantized, the MICZ-Kepler system leads to the same spectrum as
 the Coulomb problem, with a little difference consisting in the
 shift of the possible values of the orbital quantum number -- it
 starts with $|s|$. The Hamiltonian of MICZ-Kepler system which describes the
 motion of electrically charged scalar particle in the field of
 static Dirac dyon reads
 \be \label{H0}
\mathcal{H}=\frac{1}{2m} \left(\mathbf{p}-e\mathbf{A}_g
 \right)^2-\frac{e q }{r}+\frac{s^2}{2 m r^2}, \quad
\mbox{rot}\mathbf{A}_g=\frac{g \mathbf{r}}{r^3}.
 \ee
Obviously, there are many ways to construct the multi--center
generalization of the Hamiltonian \p{H0}. Of course, the
preferable generalization has to preserve the main property of the
MICZ-Kepler system - its integrability. Quite interestingly, $N=4$
supersymmetry ruled out just the unique generalization of \p{H0}.
It has been shown in \cite{kno, no} that the proper multi--center
generalization of the MICZ-Kepler system reads \be \label{H00}
\mathcal{H}=\frac{1}{2m} \left(\mathbf{p}-e \sum_{i=1}^n
\mathbf{A}_{g_i}(\mathbf{r}-\mathbf{a}_i)
  \right)^2 - e\sum_{i=1}^n
  \frac{q_i}{|\mathbf{r}-\mathbf{a}_i|}+\frac{e^2}{2
  m}\left(\sum_{i=1}^n\frac{g_i}{|\mathbf{r}-\mathbf{a}_i|}
  \right)^2, \quad
\mbox{rot}\mathbf{A}_{g_i}(\mathbf{r})=\frac{g_i \mathbf{r}}{r^3}.
\ee This Hamiltonian describes the motion of an electrically
charged scalar particle in the field of $n$ Dirac dyons sitting at
the points with coordinates $\mathbf{a}_i$. Just with such
structure of potential terms, the Hamiltonian admits $N=4$
supersymmetrization and, moreover, it describes a classically
integrable system, at least for the two centers case.

One of the possible ways to further extend the system \p{H00} is
to consider the MICZ-Kepler system on the sphere $S^3$ in the
field of $n$ Dirac dyons. Clearly, the $N=4$ supersymmetry,
provided such a superextension exists, should help to find a
proper multi--center extension. While trying to construct the
$N=4$ supersymmetric version of the MICZ-Kepler system, one may
immediately conclude that there are two possibilities to have a
sphere $S^3$ in the bosonic sector. Firstly, one may start with
the $N=4, d=1$ tensor supermultiplet \cite{tensor}, which contains
on-shell three bosonic and four fermionic components. With a
properly chosen metrics, one may get the sphere $S^3$ in the
bosonic sector. Then one may add the most general potential term,
following the general construction \cite{{IL1},{ks1}}. When
the bosonic metric is completely fixed to be the $S^3$ one, the
possible potential terms are completely defined by a function
obeying the flat three-dimensional Laplace equation. Clearly, in
such a way it is impossible to get the monopole potential on
$S^3$.

Alternatively, one may start with the $N=4, d=1$ nonlinear
supermultiplet \cite{{ikl1},{bk}},  which contains again
three bosonic and four fermionic components on-shell. After fixing
the metric, the potential terms are defined now by an arbitrary
function obeying the three-dimensional Laplace equation on $S^3$.
Just this case is what we are going to analyze in full details in
the present work. In Section 2 we shortly describe the superspace
construction of the corresponding Lagrangian and potential terms.
In Section 3 we deal with the components approach. We present the
Hamiltonian and supercharges for arbitrary potential terms. The
main properties of these potentials is that they are fully
determined by an arbitrary function which has to obey Laplace
equation on the sphere $S^3$. In Section 4 we consider two
particular cases of potential terms, i.e. with spherical and
cylindrical symmetries, which seem to be the most interesting
ones. Finally, we conclude with some comments.

\setcounter{equation}{0}
\section{$N=4$, $d=1$ nonlinear supermultiplet}
The $N=4, d=1$ nonlinear supermultiplet has been constructed in
\cite{ikl1} and then further analyzed in \cite{bk}. It is defined
in terms of the three $N=4, d=1$ superfields $\Phi, \Lambda,\bLam
$ subject to the constraints:
\bea \label{constr}
&&D^1 \Lambda=-
\Lambda D^2 \Lambda, \quad \bD_2 \Lambda= \Lambda \bD_1 \Lambda,
\quad D^2 \overline{\Lambda}=\overline{\Lambda} D^1
\overline{\Lambda}, \quad \bD_1 \overline{\Lambda}=-
\overline{\Lambda} \bD_2 \overline{\Lambda}, \quad \\ \nonumber
&&i D^1 \Phi = -D^2 \Lambda, \quad i \bD_1 \Phi = \bD_2
\overline{\Lambda}, \quad i D^2 \Phi = -D^1 \overline{\Lambda},
\quad i \bD_2 \Phi = \bD_1 \Lambda,
\eea
where spinor derivatives
are defined by \be D^i=\frac{\partial}{\partial\theta_i}+i
\bar{\theta}^i \partial_t, \quad
\bD_i=\frac{\partial}{\partial{\overline {\theta^i}}}+i \theta_i
\partial_t, \quad \{D^i,\bD_j \}=2 i \delta^i_j \partial_t. \ee
The constraints \p{constr} leave in the nonlinear supermultiplet
three physical  $\lambda, \bl, \phi$ and one auxiliary $A$ bosonic
fields and four fermionic fields $\psi_a, \overline{\psi}^a$
($a=1,2$), which may be defined as \bea \label{comp} &&\phi=\Phi|,
\quad \lambda=\Lambda|, \quad \bl=\overline{\Lambda}|, \quad
A=\left( D^1\bD_1-\bD_1 D^1 \right)\Phi|, \\ \nonumber
 &&\psi_1=\frac{1}{2}\bD_1 \Phi| \quad \psi_2=-\frac{1}{2}\bD_2 \Phi|,
\quad \overline{\psi}^1=-\frac{1}{2}D^1 \Phi|, \quad \overline{\psi}^2=\frac{1}{2}D^2 \Phi|
\eea where $|$ means $\theta_i=\overline{\theta}^j=0$. The
transformation properties of these components under $N=4$
supersymmetry read as follows: \bea \label{trans} &&\delta \lambda
=-2 i \left( \epsilon_2 - \epsilon_1 \lambda \right) \bpsi^1+2i
\left(\bar{\epsilon}^1+ \lambda \bar{\epsilon}^2 \right)\psi_2,
\quad \delta \phi =2 \left( \epsilon_1 \bpsi^1 - \epsilon_2
\bpsi^2-\bar{\epsilon}^1 \psi_1 + \bar{\epsilon} ^2 \psi_2\right),
\\ \nonumber
&&\delta\psi_1=-\frac{1}{2}\epsilon_1 \left(i
\dot{\phi}+\frac{1}{2} A \right)-\frac{1}{2}\epsilon_2 \left( 2
\dot{\bl}+4 i \psi_1\bpsi^2+i\bl\dot{\phi}+\frac{1}{2}\bl A
\right), \\ \nonumber
&&\delta\psi_2=\frac{1}{2}\epsilon_2 \left(i \dot{\phi}-\frac{1}{2} A \right)+\frac{1}{2}\epsilon_1 \left( 2 \dot{\lambda}-
4 i \psi_2\bpsi^1-i\lambda\dot{\phi}+\frac{1}{2}\lambda A \right), \\
&&\delta A=-4 i \left( \epsilon_1 \dot{\bpsi}^1+\epsilon_2
\dot{\bpsi}^2+\bar{\epsilon}^1 \dot{\psi}_1+\bar{\epsilon}^2
\dot{\psi}_2 \right) \nonumber . \eea The general sigma-model type
off-shell action has the form \cite{ikl1} \be\label{S1} S=\int d t
d \theta^2 d \bar{\theta}^2 L(\Phi, \Lambda, \overline{\Lambda}),
\ee where $L( \Phi, \Lambda, \overline{\Lambda} )$ is an arbitrary
real function of the superfields $(\Phi,\Lambda,\bLam)$. The
simplest potential term may be generated in a standard manner by
adding to the action \p{S1} the  Fayet--Iliopoulos term
\be\label{S2}
{\tilde S}_p= m \int dt A,
\ee
with $m$ being the
coupling constant. This potential term gives rise to the
interaction with the electric field, but it will never produce the
interaction with the magnetic field. Fortunately, for the
nonlinear supermultiplet there is a more general Fayet-Iliopoulos
term. Indeed, it has been shown in \cite{bk} that one may define
the generalized auxiliary component $B$ as \be B=h_\phi A +b
\dot{\lambda}+{\bar b} \dot{\bl}+a (\bpsi^1 \psi_1 - \bpsi^2
\psi_2) + a_1 \bpsi^2 \psi_1 + a_2 \bpsi^1 \psi_2, \label{B} \ee
where \bea\label{b1} && a=-8\frac{h_{\phi\phi}}{1+\lambda\bl},\;
a_1=-8ih_{\phi\bl}+8\lambda \frac{h_{\phi\phi}}{1+\lambda\bl},\;
a_2=8ih_{\phi\lambda}+8\bl \frac{h_{\phi\phi}}{1+\lambda\bl},\; \nonumber \\
&& b=2ih_\lambda +4 \bl \frac{h_\phi}{1+\lambda\bl},\;{\bar
b}=-2ih_{\bl} +4 \lambda \frac{h_\phi}{1+\lambda\bl}, \eea and $h$
obeys the Laplace equation on $S^3$: \be
h_{\phi\phi}+\left(1+\lambda \bl \right)h_{\lambda \bl}+i \lambda
h_{\lambda \phi}-i \bl h_{\bl \phi}=0. \label{lap} \ee With all
these equations \p{b1},\p{lap} being satisfied, the new auxiliary
component \p{B} transforms under $N=4$ supersymmetry through a
full time derivative \cite{bk}. Therefore, we may add to the
action \p{S1} a new generalized Fayet-Iliopoulos term:
\be\label{S3} {\hat S}= S+ m \int dt B. \ee As we will see in the
next Section, the action \p{S3} provides the most general
interaction with electric and magnetic fields.

To close this Section let us clarify in more details the differences
between linear and nonlinear $N=4$ supermultiplets. For this purpose we will
construct the most general potential term in \p{S3} for both these supermultiplets in a different way.
First of all let us rewrite the basic constraints \p{constr} as follows
\bea\label{constr_new}
&& D^1 \Lambda =i \alpha \Lambda D^1 \Phi,\; \bD_1{\overline \Lambda}=-i\alpha {\overline \Lambda}\bD_1 \Phi, \label{c_new_1} \\
&& D^2 \Lambda =-i D^1 \Phi,\; \bD_2 \Lambda =\alpha \Lambda \bD_1 \Lambda,\; iD^2\Phi=-D^1 {\overline \Lambda},\;
i\bD_2 \Phi=\bD_1 \Lambda, \label{c_new_2} \\
&& D^2 {\overline \Lambda}=\alpha {\overline \Lambda} D^1 {\overline \Lambda},\; \bD_2 {\overline \Lambda}=i \bD_1 \Phi. \nonumber
\eea
Here, we introduce the parameter $\alpha$ to discuss two cases simultaneously: with $\alpha=0$ we have the standard
linear $N=4$ tensor supermultiplet \cite{tensor}, while for the $\alpha\neq 0$ one may always rescale the superfields to achieved
$\alpha=1$ just as in the basic constraints \p{constr}. It is clear from \p{c_new_2} that the $D^2$ and $\bD_2$ derivatives from
all our superfields are expressed through $D^1$ and $\bD_1$ derivatives from the same set of superfields. This means that all
components of our (linear)nonlinear supermultiplet appear in the $N=2$ superfields ${\hat \Lambda},{\hat{\overline \Lambda}},{\hat \Phi}$
\be
{\hat \Lambda}=\Lambda_{\theta_2=\bar\theta{}^2=0},\quad
{\hat{\overline \Lambda}}=\hbL_{\theta_2=\bar\theta{}^2=0},\quad
{\hat \Phi}=\Phi_{\theta_2=\bar\theta{}^2=0},
\ee
which
depend only on $\theta_1$ and $\bar\theta{}^1$. On these $N=2$ superfields the another implicit $N=2$ supersymmetry is realized as follows
\be\label{tr11}
\delta {\hat \Lambda} = i\epsilon_2 D^1 {\hat\Phi}-\alpha \bar\epsilon{}^2 {\hat\Lambda} \bD_1 {\hat\Lambda},\;\;
\delta \hbL=-\alpha \epsilon_2 \hbL D^1 \hbL-i\bar\epsilon{}^2 \bD_1 {\hat \Phi},\;\;
\delta {\hat \Phi} =-i\epsilon_2 D^1 \hbL+i\bar\epsilon{}^2 \bD_1 \hL.
\ee
Now, one may immediately write the most general potential term as
\be\label{pot11}
S_p=m \int dt d\theta_1d\bar\theta{}^1 H({\hat \Lambda},{\hat{\overline \Lambda}},{\hat \Phi}).
\ee
where, for the time being, $H$ is an arbitrary function.

By construction, the potential term \p{pot11} is manifestly invariant
with respect to $N=2$ supersymmetry realized on the $(t,\theta_1,\bar\theta{}^1)$. With respect to implicit $N=2$ supersymmetry \p{tr11}
the integrand in \p{pot11} transforms as follows (we will write only $\epsilon_2$ part of the variation)
\be\label{tr12}
\delta H = \epsilon_2\left( H_{\hL} \delta \hL  +H_{\hbL} \delta\hbL +H_{\hat\Phi}\delta {\hat\Phi}\right) =
 -\epsilon_2\left[ -iH_\hL D^1 {\hat \Phi}+ \left( iH_{\hat\Phi}+\alpha \hbL H_\hbL\right) D^1 \hbL \right].
\ee
If we insist on the invariance of the potential term \p{pot11} under \p{tr12} the variation \p{tr12} must be represented as
\be\label{tr13}
\delta H = -\epsilon_2 D^1 G({\hat \Lambda},{\hat{\overline \Lambda}})=
 -\epsilon_2\left[ \left( G_{\hat\Phi}+i\alpha\hL G_{\hL}\right)D^1{\hat\Phi} + G_\hbL D^1 \hbL\right],
\ee
where $G({\hat \Lambda},{\hat{\overline \Lambda}},{\hat\Phi})$ is an arbitrary function on its arguments and we used the constraints \p{c_new_1}.
Comparing \p{tr12} and \p{tr13} we will get the following conditions
\be\label{int1}
iH_{\hat\Phi}+\alpha \hbL H_\hbL = G_\hbL, \quad -i H_\hL= G_{\hat\Phi}+i\alpha \hL G_\hL.
\ee
The integrability of the constraints \p{int1} gives us the desired constraints on the super-potential $H({\hat \Lambda},{\hat{\overline \Lambda}},{\hat\Phi})$
\be\label{Lap11}
\left(1+\alpha^2 \hL \hbL\right) H_{\hL\hbL}+H_{{\hat\Phi}{\hat\Phi}}+i\alpha\left(\hL H_{{\hat\Phi}\hL}-\hbL H_{{\hat\Phi}\hbL}\right)=0.
\ee
Thus we conclude, the potential term \p{pot11} is invariant with respect to $N=4$ supersymmetry if its integrand obeys to the equation \p{Lap11}.

Now the differences between liner and nonlinear supermultiplet becomes transparent: the potential term for the nonlinear supermultiplet is
defined by a harmonic on $S^3$ super function, while for the linear tensor supermultiplet this function has to obey flat Laplace equation $(\alpha=0)$.
Being rewritten in the components, the potential term \p{pot11} is coincides with the potential in \p{S3} after identification
\be
H({\hat \Lambda},{\hat{\overline \Lambda}},{\hat\Phi})|_{\theta_1=\bar\theta{}^1=0}=h(\lambda,\bar\lambda,\phi).
\ee

It is worth to note that the most general $N=4$ supersymmetric
action for the conformally flat case has been constructed many
years ago in \cite{Sm}. We would like to stress again that while
the kinetic parts in the $N=4$ actions for linear and nonlinear
supermultiplet describe the conformally flat three-dimensional
bosonic manifold, the structure of the potential terms is
completely different in these cases. The main reason for this is
the nonlinear realization of the off-shell supersymmetry on the
components in the nonlinear case \p{trans}. This is the reason why
the action \p{S3} cannot be obtained within the approaches in
\cite{tensor}, \cite{Sm}.

Moreover, in the next Section we will explicitly demonstrate that even the kinetic parts of the actions
are different for the linear and nonlinear supermultiplets.

\setcounter{equation}0
\section{Components description: Lagrangian and Hamiltonian}
In order to clarify the structure of the action \p{S3}, let us go
to components. For doing this, one should perform an integration
over Grassmann variables in \p{S3} (with the constraints
\p{constr} imposed), and then eliminate the auxiliary component
$A$. Before carrying out this task, let us make two essential
comments.

First of all, we are interested to get a $S^3$ sphere in the
bosonic sector of the action. It has been shown in \cite{ikl1}
that for this case the superfield Lagrangian $L$ in \p{S1} has to
be chosen as \be\label{L} L=\ln(1+\Lambda\bLam). \ee Secondly,
after going to components, the kinetic terms for the fermions read
\be\label{Lf} L_f=\frac{8i}{1+\lambda\bl}\left[ \dot{\psi}_1
\bpsi^1 +\dot{\psi}_2 \bpsi^2+ \frac{1}{1+\lambda\bl}\left(
\dot\lambda \psi_1\bpsi^2 -\dot{\bl} \psi_2\bpsi^1 -
\lambda\dot{\bl}\psi_1\bpsi^1- \bl\dot{\lambda}\psi_2\bpsi^2
\right) \right]. \ee One may easily check that this expression can
be drastically simplified after passing to the new fermionic
fields\footnote{The same transformations have been used in
\cite{ikl1} for the case of a particle on $S^2$.}
\be\label{newferm} \psi=\frac{\bpsi^2+\bl \bpsi^1}{1+\lambda\bl},
\quad \xi=\frac{\bpsi^1-\lambda \bpsi^2}{1+\lambda\bl}, \ee in
term of which it take the standard free form \be\label{Lf1} L_f
=-8i\left( \psi\dot{\bpsi}+\xi\dot{\bxi}\right). \ee

Taking all this into account, we may perform the integration over
Grassmann variables and eliminate the auxiliary component $A$.
After passing to the newly defined fermions \p{newferm}, we end up
with the following action: \bea \label{Sfin} S&=&\int dt\left[
\frac{4 \dot{\lambda} \dot{\bl}}{\left(1+\lambda \bl \right)^2}+
      \left( \dot{\phi}+ i \frac{\dot{\lambda}\bl - \dot{\bl}\lambda}{1+\lambda \bl} \right)^2
- m^2 h_{\phi}^2 + 2 m h_{\phi}\frac{\partial_t\left(\lambda\bl\right)}{1+\lambda\bl}+
2 i m \left( h_{\lambda}\dot{\lambda}-h_{\bl}\dot{\bl}\right) \right.\nonumber \\
&&\left. -8i\left( \psi\dot{\bpsi}+\xi\dot{\bxi}\right) -8m \left(
1 + \lambda \bl \right) \left[ h_{\lambda \bl}\left(\bxi \xi -
\bpsi \psi \right)+\left(i h_{\phi\lambda}+ \bl h_{\lambda \bl}
\right)\bpsi \xi+\left( -i h_{\phi \bl}+\lambda h_{\lambda \bl}
\right)\bxi \psi \right] \right]. \eea The bosonic kinetic terms
of the action \p{Sfin} describe just the sphere $S^3$ in
stereographic coordinates. What is a really interesting is that
the $N=4$ supersymmetrization of this $S^3$ can be achieved by
adding four free fermions. Let us remind, that just the same
phenomenon appears in the case of the $N=4$ supersymmetrization of
the sphere $S^2$ \cite{ikl1}. In addition, in the action \p{Sfin}
there are potential terms which are completely specified by the
function $h$ obeying Laplace equation on $S^3$ \p{lap}.

Before going to the construction of the Hamiltonian and
supercharges, let us note that the kinetic part of the action
\p{Sfin} can be brought into the simpler form \be S_{kin}=\int dt
\left(\frac{4\left(\dot{\mathbf{x}}\dot{\mathbf{x}}
\right)}{\left(1+\mathbf{x}^2 \right)^2} -
 8 i \left(\psi \dot{\bpsi}+\xi\dot{\bxi} \right) \right), \label{LB}
\ee
where the new coordinates $\mathbf{x}=(x_1, x_2, x_3)$ are related with the
initial ones as
\bea \lambda=\frac{2x_3+ i (
1-\mathbf{x}^2)}{2 (x_1+ i x_2)}, \quad \bl=\frac{2  x_3- i (
1-\mathbf{x}^2)}{2 (x_1 -i x_2)}, \quad e^{i \phi}=-\frac{x_1-i
x_2}{x_1+i x_2}. \label{cfc}
\eea

The action \p{LB} yields a perfect opportunity to further clarify
the differences between linear and nonlinear supermultiplets. From
the paper \cite{Sm} we know that the $N=4$ supersymmetric action
with linear supermultiplet has the four-fermionic term \be S\sim
\int dt \left[ G \dot{\mathbf{v}} \dot{\mathbf{v}}- \left(
\triangle G -\frac{ \partial_m G \partial_m
G}{2G}\right)\psi\bpsi\xi\bxi + L_{fer}\right]. \ee Here,
$G(v^m),\; m=1,2,3$ is an arbitrary metric and $L_{fer}$ stands
for the terms which are quadratic in fermions. Clearly, for the
sphere $S^3$ this four fermionic term unavoidably appears in the
action. In the same time, the action \p{LB}, being $N=4$
supersymmetric, does not contain such term. Thus, the same bosonic
manifold, the sphere $S^3$ in our explicit example, can be
supersymmetrized in two different ways. The reason is the
existence of two {\it different} off-shell realizations of $N=4$
supersymmetry on the three physical bosons, four fermions and one
auxiliary field. Thus, the $N=4$ mechanics we are considering here
is different from those one constructed in \cite{Sm}.

Due to the extremely simple structure of the action \p{Sfin}, the
construction of the Hamiltonian does not contain any
peculiarities. As usual, one should define the momenta $p_\lambda,
p_{\bl}, p_\phi, \pi_\psi,\pi_\xi$ \bea\label{pb1} && p_\lambda=
\frac{4\dot{\bl}}{(1+\lambda\bl)^2}+ 2i
\frac{\bl}{1+\lambda\bl}\left( \dot{\phi}+
i\frac{\dot{\lambda}\bl-\lambda\dot{\bl}}{1+\lambda\bl}\right)
+2imh_\lambda +2mh_\phi \frac{\bl}{1+\lambda\bl},\nonumber \\
&& p_\phi=2\left( \dot{\phi}+
i\frac{\dot{\lambda}\bl-\lambda\dot{\bl}}{1+\lambda\bl}\right),
\quad \pi_\psi= 4i \bpsi,\; \pi_\xi=4i \bxi, \eea and introduce
the canonical Poisson brackets \be\label{pb2} \left\{\lambda,
p_\lambda\right\}=\left\{\phi,p_\phi\right\}=1,\quad \left\{
\psi,\pi_\psi \right\}=\left\{ \xi,\pi_\xi \right\}=-1. \ee {}From
the explicit form of the fermionic momenta \p{pb2} it follows that
we have second-class constraints. In order to resolve them, we
will pass to the Dirac brackets for the canonical
variables\footnote{From now on, the symbol $\left\{,\right\}$
stands for the Dirac brackets.}
\bea \label{pb3} &&\{\lambda,
\tilde{p}_{\lambda} \}=1, \quad \{\bl, \tilde{p}_{\bl} \}=1, \quad
\{\psi,\bpsi\}=\frac{i}{8}, \quad \{\xi, \bxi\}=\frac{i}{8} \\
\nonumber &&\{ p_{\phi}, \tilde{p}_{\lambda}\}=2 m
h_{\phi\phi}\frac{\bl}{1+\lambda \bl}+2 i m h_{\phi \lambda}, \;
\{ p_{\phi}, \tilde{p}_{\bl}\}=2 m
h_{\phi\phi}\frac{\lambda}{1+\lambda \bl}-2 i m h_{\phi \bl}, \\
\nonumber &&\{\tilde{p}_{\lambda},\tilde{p}_{\bl} \}= - 2 i m
\left( h_{\lambda \bl}-\frac{h_{\phi \phi}}{1+\lambda \bl}
\right),
\eea
where the bosonic momenta
$(\tilde{p}_{\lambda},\tilde{p}_{\bl})$ have been defined as
\bea\label{pp}
&&\tilde{p}_{\lambda}=p_{\lambda}-m A_{\lambda}, \quad
A_{\lambda}=2 h_{\phi}\frac{\bl}{1+\lambda \bl}+2 i h_{\lambda},
\\ \nonumber &&\tilde{p}_{\bl}=p_{\bl}-m A_{\bl}, \quad A_{\bl}=2
h_{\phi}\frac{\lambda}{1+\lambda \bl}-2 i h_{\bl}
\eea
Now, one
may check that the following supercharges:
\bea \label{QQ}
&&Q_1=(\tilde{p}_{\phi}+i \lambda \tilde{p}_{\lambda})(\xi+\lambda
\psi)+ i \tilde{p}_{\bl} (\psi - \bl \xi)+8\psi \bpsi \xi + 2 i m
h_{\phi}\xi, \\ \nonumber &&Q_2=(\tilde{p}_{\phi}-i \bl
\tilde{p}_{\bl})(\psi-\bl \xi)+ i \tilde{p}_{\lambda} (\xi +
\lambda \psi)-8\psi \xi \bxi - 2 i m  h_{\phi}\psi, \\ \nonumber
&&\bQ^1=(\tilde{p}_{\phi}-i \bl \tilde{p}_{\bl})(\bxi+\bl \bpsi)-
i \tilde{p}_{\lambda} (\bpsi - \lambda \bxi)+8\psi \bpsi \bxi -2 i
m  h_{\phi}\bxi, \\ \nonumber
&&\bQ^2=(\tilde{p}_{\phi}+i \lambda
\tilde{p}_{\lambda})(\bpsi-\lambda \bxi)- i \tilde{p}_{\bl} (\bxi
+ \bl \bpsi)-8\bpsi \xi \bxi + 2 i m  h_{\phi}\bpsi \eea
and the Hamiltonian
\bea \label{ham} H&=&\frac{(1+\lambda\bl)^2}{4} \left(
\tilde{p}_\lambda -i \frac{\bl p_\phi}{1+\lambda\bl} \right)\left(
\tilde{p}_{\bl} + i \frac{\lambda p_\phi}{1+\lambda\bl} \right)
+\frac{1}{4}p_\phi^2 + m^2 h_\phi^2 \nonumber \\
&&+8m \left( 1 + \lambda \bl \right) \left[h_{\lambda
\bl}\left(\bxi \xi - \bpsi \psi \right)+\left(i
h_{\phi\lambda}+\bl h_{\lambda \bl} \right)\bpsi \xi+ \left( -i
h_{\phi \bl}+\lambda h_{\lambda \bl} \right)\bxi \psi \right]
\eea
form the standard $N=4$ superalgebra
\be
\left\{ Q_i, \bQ^j
\right\} =\frac{i}{2}\delta^j_iH,\quad \left\{ Q_i, Q_j \right\}=\left\{
\bQ^i, \bQ^j \right\} =0. \ee

With this, we completed the classical description of $N=4$
supersymmetric mechanics on the sphere $S^3$. The corresponding
Hamiltonian and supercharges are defined by \p{QQ} and \p{ham}.
The freedom to choose the proper potential terms is hidden in one
arbitrary function $h$ obeying the Laplace equation on the $S^3$
\p{lap}. Next, we analyze some specific interesting cases for the
potential terms.

\setcounter{equation}0
\section{Potentials}
The potential terms in the Hamiltonian \p{ham} are completely
defined by the function $h$ obeying  (\ref{lap}). Clearly, the
most interesting potentials have to possess some additional
symmetries. In this respect, the spherical symmetry of the
solution seems to be the most important case. Let us firstly
consider just such a type of potential.

\subsection{Spherically symmetric potential}
The spherical symmetry, being rather hidden in stereographic
coordinates, is quite evident in conformally flat coordinates
\p{cfc}. Remembering the relations between the stereographic
coordinates $\lambda, \bl, \phi$ and the conformally flat ones
\p{cfc}, one may easily find that the spherically symmetric case
corresponds to a function $h$ which depends only on the radius of
$S^3$ - the coordinate $y$:
\be
y=\frac{e^{i\frac{\phi}{2}}\bl +
e^{-i{\frac{\phi}{2}}}\lambda }{\sqrt{1+\lambda \bl}}=
2\frac{\mathbf{x}^2-1}{\mathbf{x}^2+1}. \label{y1}
\ee
Let us
remind that the potential term is defined in terms of $h_\phi$,
which also obeys the Laplace equation \p{lap}. This means that we
have to pick up for $h$ that solution which will give us the
spherically symmetric $h_\phi$. It is rather easy to find that the
proper solution is \be h_{\phi}=a-2b\frac{y}{\sqrt{4-y^2}},
\label{hph1} \ee where $a$ and $b$ are arbitrary constants. The
other derivatives of the function $h$ which appear in  the
supersymmetric Hamiltonian (\ref{ham}) and Dirac brackets \p{pb3}
are
\bea \label{Hder} &&h_{\phi\phi}=\frac{4 i b e^{i
\frac{\phi}{2}}\left(1+\lambda \bl \right)\left(\bl - e^{-i \phi}
\lambda\right)}{\left(4 - e^{- i \phi} \left(\lambda - e^{i \phi}
\bl \right)^2 \right)^{3/2}}, \; h_{\lambda \bl}=\frac{4 i b e^{i
\frac{\phi}{2}} \left(\bl - e^{-i \phi} \lambda \right)}
{\left(4 - e^{- i \phi} \left(\lambda - e^{i \phi} \bl \right)^2 \right)^{3/2}}, \nonumber \\
\\ \nonumber
&&h_{\phi\lambda}=\frac{4 b e^{-i
\frac{\phi}{2}}\left(2+\bl\left(\lambda - e^{i \phi}
\bl\right)\right)}{\left(4 - e^{- i \phi} \left(\lambda - e^{i
\phi} \bl \right)^2 \right)^{3/2}}, \; h_{\phi\bl}=\frac{4 b e^{i
\frac{\phi}{2}}\left(2+\lambda\left(\bl - e^{-i \phi}
\lambda\right)\right)}{\left(4 - e^{- i \phi} \left(\lambda - e^{i
\phi} \bl \right)^2 \right)^{3/2}}.
\eea
When rewritten in
conformally flat coordinates, these expressions read
\bea\label{Hder2}
&& h_{\phi}=a - b
\frac{1-\mathbf{x}^2}{|\mathbf{x}|},\;h_{\phi\phi}=-\frac{b
(1+\mathbf{x}^2)^2 x_3}{4 |\mathbf{x}|^3},\; h_{\lambda
\bl}=-\frac{b (x_1^2+x_2^2)x_3}{|\mathbf{x}|^3}, \\ \nonumber &&
h_{\phi \lambda}=-i \frac{b (x_1+i x_2)(2 \mathbf{x}^2-i
(1-\mathbf{x}^2)x_3)}{2 |\mathbf{x}|^3}, \;
 h_{\phi \bl}=i \frac{b (x_1-i x_2)(2 \mathbf{x}^2+i (1-\mathbf{x}^2)x_3)}{2 |\mathbf{x}|^3}.
\eea
Therefore, the Hamiltonian (\ref{ham}) in the case of
spherically symmetric potentials reads
\bea \label{MICZ1}
\mathcal{H}&=&\frac{(1+\lambda\bl)^2}{4} \left(
\tilde{p}_\lambda -i \frac{\bl p_\phi}{1+\lambda\bl} \right)\left(
\tilde{p}_{\bl} + i \frac{\lambda p_\phi}{1+\lambda\bl} \right)
+\frac{1}{4}p_\phi^2+m^2 \left(a -
b\frac{1-\mathbf{x}^2}{|\mathbf{x}|} \right)^2+ 2m b
(1+\mathbf{x}^2)^2\frac{\mathbf{x}}{|\mathbf{x}|^3}(\bar{\chi}\mathbf{\sigma}\chi)= \nn
&&
\frac{(1+\mathbf{x}^2)^2}{4}\left( \mathbf{p}-\mathbf{A}\right)^2+m^2 \left(a -
b\frac{1-\mathbf{x}^2}{|\mathbf{x}|} \right)^2+ 2m b
(1+\mathbf{x}^2)^2\frac{\mathbf{x}}{|\mathbf{x}|^3}(\bar{\chi}\mathbf{\sigma}\chi),
\eea
where we combined the fermions $\psi, \bpsi, \xi, \bxi$ into
the $SU(2)$ spinor
$ \chi=\left( \begin{array}{lcr}
\psi \\
\xi
\end{array}
\right), $ with  $\sigma_i$, $i = 1,2,3$ being Pauli matrices. Let
us stress that the Hamiltonian \p{MICZ1} is just a particular case
of the Hamiltonian \p{ham}, when the potential is chosen to be
spherically symmetric and we partly use the coordinates \p{cfc}.
Therefore, it also appears in the anticommutators of the
supercharges \p{QQ}, as it occurs also for the Hamiltonian
\p{ham}.

As it was argued in \cite{no}, the Hamiltonian of the MICZ--Kepler
system on an arbitrary three-dimensional space with
$so(3)$-invariant conformally flat metric
$ds^2=G(r)\left(dx_1^2+dx_2^2+dx_3^2 \right)$ should have the form
\be \label{hams1}  \mathcal{H}=\frac{1}{2 G(r)}
\left(\mathbf{p}-e\mathbf{A}_g
 \right)^2+\frac{e^2 (g\phi)^2}{2}-e q \phi, \quad
\mbox{rot}\mathbf{A}_g=-g \mbox{ grad}\phi,
\ee
where the Coulomb
potential $\phi$, which is the $so(3)$-invariant solution of the
Laplace equation
 \begin{eqnarray} \label{laps}
 \frac{\partial}{\partial
x^i}\left(G^{1/2}\frac{x^i}{r}\frac{d \phi}{d r}\right)=0,
\end{eqnarray}
reads \be \label{sollap} \phi=a+b\int\frac{d r}{ r^2 \sqrt{G(r)}},
 \ee
with $a$ and $b$ denoting arbitrary constants. In the case of the
sphere $S^3$ with $G(r)=\frac{4}{(1+r^2)^2}$, one may immediately
conclude that the Coulomb potential has the form given by the
first equation in (\ref{Hder2}). Thus, the bosonic part of the
Hamiltonian (\ref{MICZ1}) does completely coincide with the
Hamiltonian of the charged particle on the sphere $S^3$ moving in
the field of Dirac dyon, whereas the fermionic part is just the
Zeeman  energy, $U_Z=-\mathbf{B}\mathbf{M}$, i.e.
the energy of the interaction between the particle magnetic moment
$\mathbf{M}=8e\left( \bar{\chi}
\sigma \chi \right)$
and the magnetic field of the dyon, which has the monopole-like
nature $\mathbf{B}= g\frac{1}{G(x)}\frac{
\mathbf{x}}{x^3}=g \frac{\left(1+\mathbf{x}^2 \right)^2}{4}\frac{
\mathbf{x}}{x^3}$. Thus, one should identify $b$ with magnetic
charge of the dyon $g$ and $m$ - with the electric charge of the
moving particle $e$. Moreover, in order to obtain proper Coulomb
potential term corresponding to the interaction between moving
particle and electric charge of the dyon $e q \phi$ one should put
$a=\frac{e q}{2 g}$: \bea \label{MICZ_f}
&&\mathbf{H}=\frac{\left(1+\mathbf{x}^2 \right)^2
}{4}\mathbf{p}^2+e^2 \left(a+g \phi
\right)^2+\mathbf{B}\mathbf{M},  \\ \nonumber
&&\phi=\frac{1-\mathbf{x}^2}{x}, \quad \mathbf{B}=g
\frac{\left(1+\mathbf{x}^2 \right)^2}{4}\frac{ \mathbf{x}}{x^3}
 \eea

Thus, we conclude that the Hamiltonian \p{MICZ1} describes the
$N=4$ supersymmetric MICZ--Kepler system on $S^3$. Of course, one
may include into the Hamiltonian an arbitrary number of monopoles
\p{Hder2}, in full analogy with the flat case \cite{kno}. We would
like to stress that, while in the fermionic sector all terms
coming from different monopoles will just sum up, the
corresponding bosonic potential will be the square of the sum. So,
additional cross-terms will appear. These terms are definitely
needed, in order to have $N=4$ supersymmetry. Moreover, in a full
analogy with the flat case, just this structure of the potential
seems to be absolutely necessary for the integrability of the
model, at least in the two monopoles case.

\subsection{Cylindrically symmetric potential}
It is clear that the  stereographic coordinates are not so
suitable to describe the spherically symmetric solutions  of the
Laplace equation on $S^3$. The ``radial'' variables $y$ \p{y1}
look rather artificial in stereographic coordinates. Moreover,
when analyzing the structure of \p{y1} one may wonder whether the
similar combination $y_3$
\be
y_3=i\frac{e^{i\frac{\phi}{2}}\bl -
e^{-i{\frac{\phi}{2}}}\lambda }{\sqrt{1+\lambda \bl}}, \label{y2}
\ee
is suitable to get the particular solution of the Laplace
equation. Indeed, it turns out that this is precisely the case.
The corresponding solution has the same form as \p{hph1}:
\be
{\tilde h}_{\phi}=a_1-2b_1\frac{y_3}{\sqrt{4-y_3^2}}. \label{hph3}
\ee
Passing to conformally flat coordinates, we get
\be\label{cyl}
{\tilde h}_\phi= a_1 - \frac{2 b_1 x_3}{\sqrt{(1+\mathbf{x}^2)^2-4
x_3^2}}.
\ee
The remaining needed functions appearing in the
Hamiltonian can be easily found from \p{hph3}. At any rate, the
explicit form of the potential \p{cyl} yields us informations
about the cylindrical symmetry (for rotations around the $x_3$
axis) of the solution.

It is worth to notice that the similar cylindrically symmetric
solutions, with $x_3$ being replaced by $x_1$ and $x_2$, follow {}
from two other solutions of the Laplace equations. They have the
same form as \p{hph3}, with the replacements $y_3 \rightarrow y_1$
and $y_3 \rightarrow y_2$, where
\be\label{yy}
y_2=\frac{e^{i\frac{\phi}{2}}+ e^{-i{\frac{\phi}{2}}}
}{\sqrt{1+\lambda \bl}}, \quad y_1=i\frac{e^{i\frac{\phi}{2}}-
e^{-i{\frac{\phi}{2}}} }{\sqrt{1+\lambda \bl}}. \ee

Finally, let us note that one may freely combine an arbitrary
number of spherically symmetric monopoles with an arbitrary number
of cylindrically symmetric ones, situated at arbitrary points.
Moreover, as it is completely clear from the form of the $S^3$
Laplace in stereographic coordinates \p{lap}, one may generate a
new solution from the known ones by differentiating/integrating
the latter over $\phi$. In this way one may produce a series of
solutions originating from spherical/cylindrical symmetric
monopoles. Of course, in order to decide which ones among them are
really interesting, one should involve either physical arguments
or integrability properties.

\setcounter{equation}{0}
\section{Conclusion}
In this paper we derived the Hamiltonian and supercharges of the
$N=4$ supersymmetric MICZ-Kepler system on $S^3$. We found the
proper potential terms with spherical and cylindrical symmetry. In
the case of spherically symmetric potential, we explicitly showed
that in the bosonic sector our Hamiltonian describes the motion of
the probe particle on the sphere $S^3$ in the field of $n$ Dirac
dyons sitting at arbitrary points. The structure of the potential
terms in the the multi--center cases is very similar to the
``flat'' MICZ-Kepler system \cite{kno}. It is quite important
that, while in the fermionic sector all terms coming from
different monopoles will just be summing up, the corresponding
bosonic potential will be the square of the sum. So, additional
cross-terms will appear. These cross-terms are quite necessary for
having $N=4$ supersymmetry.

One of the most interesting immediate problems is to analyze the
integrability properties of the constructed system. We expect
that, at least the two dyons system, will correspond to an
integrable case. Another intriguing question concerns the
integrability of the cylindrically symmetric potentials. Finally,
the very simple structure of the $N=4$ supersymmetrization of the
particle on $S^3$ raises the question of the existence of its
$N=8$ superextensions. Unfortunately, at present, no known example
exists for $N=8$ supersymmetric systems on constant curvature
bosonic manifolds. Our results, presented in this work, show that
the relevant $N=8$ supermultiplet, if it exists, should correspond
to some extension of the nonlinear $N=4$ supermultiplet. The
corresponding construction is rather involved. Moreover, the
structure of the possible potential terms is much more restricted
in the case of $N=8$ supersymmetry. We are hoping to report the
corresponding results elsewhere.

Finally, we would like to comment the question raised in \cite{no}: whether it is possible to construct
$N=4$ supersymmetric mechanics in which function describing the potential term obeys the same
equation as metrics in the bosonic kinetic terms did. In the present paper we demonstrated that
such situation indeed realized in the case of the sphere $S^3$. But the main ingredient we used was the
nonlinear $N=4$ supermultiplet intrinsically related with $S^3$ \cite{ikl1}. Now we do not know another
$N=4$ supermultiplets with three physical bosonic components, beside linear tensor and nonlinear ones.
So, the construction of a such $N=4$
supersymmetric mechanics seems to be a rather problematic.

\section{Acknowledgements}
We are indebted to Armen Nersessian for valuable discussions.

S.K. and V.O. thank the INFN-Laboratori Nazionali di Frascati,
where this work was completed, for warm hospitality. This work was
partly supported by grants RFBR-06-02-16684, 06-01-00627-a, DFG
436 Rus 113/669/03 and by INTAS under contract 05--7928.

\end{document}